\documentclass[twocolumn,prl,superscriptaddress,longbibliography]{revtex4-1}
\usepackage[utf8]{inputenc}
\usepackage{amsmath,amsfonts,amssymb}
\usepackage{graphicx}

\usepackage{hyperref}
\usepackage[usenames,dvipsnames]{xcolor}

\usepackage{tikz}
\hypersetup{colorlinks=true, linkcolor=BrickRed, urlcolor=blue!50!black, citecolor=blue!50!black}

\newcommand\diff{\mathrm{d}}

\renewcommand{\vec}[1]{\mathbf{#1}}
\renewcommand{\phi}[0]{\varphi}

\usepackage[normalem]{ulem}

\begin{document}
\title{Motility-induced temperature difference in coexisting phases}
\date{\today}
\def\hhu{\affiliation{Institut f\"ur Theoretische Physik II: Weiche Materie, \\
  Heinrich-Heine Universit\"at D\"usseldorf,
  Universit\"atsstr.~1, 40225 D\"usseldorf, Germany}}
\def\inn{\affiliation{Institut f\"ur Theoretische Physik,
  Universit\"at Innsbruck, Austria}}
\def\tud{\affiliation{Theorie Weicher Materie, Fachbereich Physik, Technische Universität Darmstadt, Hochschulstraße 12, 64289 Darmstadt, Germany}}


\author{Suvendu Mandal}
\email{mandal@hhu.de}
\affiliation{Institut f\"{u}r Theoretische Physik II: Weiche Materie, Heinrich-Heine-Universit\"{a}t D\"{u}sseldorf, D-40225 D\"{u}sseldorf, Germany}
\author{Benno Liebchen}
\email{liebchen@fkp.tu-darmstadt.de}
\affiliation{Institut f\"{u}r Theoretische Physik II: Weiche Materie, Heinrich-Heine-Universit\"{a}t D\"{u}sseldorf, D-40225 D\"{u}sseldorf, Germany}
\affiliation{Theorie Weicher Materie, Fachbereich Physik, Technische Universität Darmstadt, Hochschulstraße 12, 64289 Darmstadt, Germany}
\author{Hartmut L\"owen}
\email{hlowen@hhu.de}
\affiliation{Institut f\"{u}r Theoretische Physik II: Weiche Materie, Heinrich-Heine-Universit\"{a}t D\"{u}sseldorf, D-40225 D\"{u}sseldorf, Germany}

\begin{abstract}
Unlike in 
thermodynamic equilibrium where coexisting phases always have the same temperature, 
here we show that systems comprising ``active'' self-propelled particles can self-organize into 
two coexisting phases at different kinetic temperatures, which are separated from each other 
by a sharp and persistent temperature gradient. 
Contrasting previous studies which have focused on overdamped descriptions of active 
particles, 
we show that a ``hot-cold-coexistence'' occurs if and only if accounting for inertia, which is significant in a broad range of systems such as activated dusty plasmas, microflyers, whirling fruits or beetles at interfaces. 
Our results exemplify a route to use active particles to create a self-sustained temperature gradient across coexisting phases, a phenomenon, which is fundamentally beyond 
equilibrium physics. 
\end{abstract}

\maketitle

\paragraph*{Introduction.--}
In equilibrium systems, entropy maximization (or free energy minimization) requires thermal,
mechanical and chemical equilibrium among coexisting
phases. Conversely, in nonequilibrium no fundamental
law forbids different temperatures in coexisting phases,
evoking the question if a specific mechanism exists which
can generate such a difference. Such a mechanism may
appear counterintuitive, as heat-gradients, unless they are sustained by a localized heat-source such as a star performing nuclear fusion, 
usually cause processes opposing
them and driving the system towards thermal equilibrium (unless for ideal isolation): For example, a temperature difference in the air evokes a balancing wind, and 
air friction cools down a radiator once switched off. 

Here we report and systematically explore a surprisingly different scenario, where particles self-organize into coexisting phases sustaining 
different temperatures. This two temperature coexistence occurs spontaneously in a uniform system and 
remarkably, there is no heat flux at steady state, because the gradient in kinetic temperature is balanced by a self-sustained, opposite density gradient. 
A ``hot'' and a ``cold'' phase are allowed to coexist in principle, as the system we consider comprises
self-propelled microparticles which allow the system to 
bypass equilibrium thermodynamics.   

By now, we know that such microparticles, often described 
as ``active Brownian particles''~\cite{Romanczuk2012,Cates2015,Ni:2013,Kurzthaler:2018,Winkler:2015virial},
can self-organize into a liquid phase, coexisting with a gas-phase, even 
when interacting purely repulsively~\cite{Tailleur2008,Fily2012,Patch:2018curvature,Buttinoni2013,Stenhammar2013,Stenhammar:2014soft,Redner2013,Digregorio:2018,Mokhtari:2017,Solon2018generalized,Levis:2017active,Siebert:2018}. Coined as ``motility-induced phase separation'', or MIPS, this phenomenon has advanced to 
a key paradigm in the physics of self-propelled particles.
When the microparticles are overdamped, like microorganisms in a solvent~\cite{Elgeti2015} or active colloidal microswimmers~\cite{Bechinger2016,Huang:2019,Aubret:2017,Aubret:2018b,Aubret:2018b,Maggi:2017}, 
they are equally fast in both phases. Hence, despite the presence of active microparticles, liquid and gas as emerging from MIPS have 
identical kinetic temperatures, just like for liquid-gas phase separation in equilibrium. 
(Note that MIPS involves a slow-down of particles in regions of high density~\cite{Tailleur2008,Cates2015}; which occurs however only for the 'coarse grained self-propulsion', not for the actual velocity determining the kinetic temperature, as further discussed below.)

When releasing the overdamped standard approximation, as relevant e.g. for beetles at interfaces~\cite{Mukundarajan:2016}, whirling fruits~\cite{Rabault:2019} 
microflyers~\cite{Scholz2018b} or activated dusty plasmas~\cite{Morfill:2009}, both the phase diagram and the properties of the 
contained phases change dramatically, as we show in this Letter. 
In particular, while MIPS generally requires a sufficiently large self-propulsion speed $v_0$ to occur, specifically for underdamped particles it breaks down again if $v_0$ is too large, i.e. MIPS is reentrant in the presence of inertia~\cite{Suma:2014}. This is because MIPS also requires particles to slow-down (regarding their directed motion) in regions of high density~\cite{Cates2015}: such a slow-down occurs instantaneously upon collisions of overdamped particles, but in the presence of inertia, particles bounce back from collisions and do not slow down much before experiencing subsequent collisions. Thus, at very large $v_0$, underdamped particles can exchange their kinetic energies before slowing down much and MIPS breaks down.



\begin{figure}
\includegraphics[width=\linewidth]{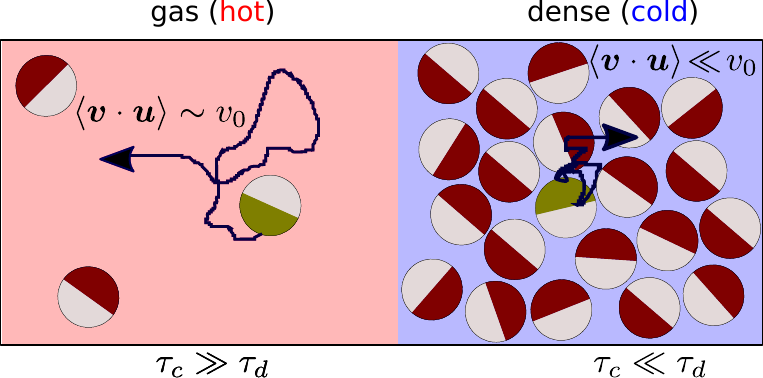}
\caption{\label{fig:hot_and_cold_cartoon}
Scheme of the phase-separated state associated with a hot-cold coexistence in underdamped active particles. Particles self-propel with the colored cap ahead (brown; greenish for the tagged particle). Active particles move with $\sim v_0$ in the gas phase, but can be an order of magnitude slower in the dense phase.}
\end{figure}

To see which physical mechanism controls the kinetic temperature difference (to be distinguished from the effective temperature \cite{Cugliandolo:2011effective,Nardini:2017,Levis:2015single,Preisler:2016}) in coexisting phases, 
consider the collision of an active underdamped particle moving with a fixed orientation towards an 
elastically reflecting wall.
This problem is equivalent to a bouncing ball experiencing friction and gravity (see Supplemental Material for details): 
while reaching a terminal speed ($v_0$) when falling in free space, the ball continuously slows down, when reflected by a wall, even when the 
collisions are elastic. 
Analogously, particles essentially move with $v_0$ in the gas phase, where they rarely collide, 
but slow down when entering the dense liquid phase, due to successive collisions with other particles (see Fig.~\ref{fig:hot_and_cold_cartoon}). 
Notice that 
inelastic collisions among the particles provide an alternative, but mechanistically unrelated, 
route to achieve a remarkable hot-cold coexistence, 
which has been discussed for vibrated granular particles, where particles dissipate energy due to inelastic collisions~\cite{Yuta:2015,Roeller:2011,Schindler:2018}. In contrast, 
for the microparticles we consider, no inelastic collisions are required: the emergence of coexisting temperatures is based on the interplay of 
activity and weak inertia.

Our results exemplify a generic route to use active particles to create a self-sustained temperature gradient across coexisting phases, 
a phenomenon, which is fundamentally beyond 
equilibrium physics. This contrasts the overdamped standard case, 
which has been predominantly explored in active matter physics so far and leads to 
a dynamics which can be essentially mapped onto an equilibrium system at a coarse grained level~\cite{Tailleur2008,Cates2015} 
yielding a phase transition which is consistent with an equilibrium liquid-gas transition~\cite{Levis2017}.
Thus, the existence of temperature differences in coexisting phases indicates a change of the nature of MIPS, 
when releasing the overdamped standard approximation: it changes from a liquid-gas like transition 
to a new type of phase transition having no counterpart in 
equilibrium. 
Accordingly, part of phenomenology of MIPS~\cite{Tailleur2008,Krinninger:2016,Solon:2018,Speck:2014,Wysocki:2014}, 
a key result in active matter physics, 
is even broader than anticipated previously - but was curtained by the overdamped standard approximation in previous studies. 
 
\begin{table}[t]
        \centering
        {
                \begin{tabular}{c | c }
                        \hline
                        \hline\\[-8pt]
                        persistence time \quad \quad \quad \quad \quad  \quad \hspace{0.5em} & $\tau_p=1/D_r$\\[8pt]
                        mean time between collisions \hspace{0.5em} &\quad \quad $\tau_c=\pi \sigma/(4v_0 \varphi)$\\[8pt]
                        inertial time \quad \quad \quad \quad \quad \quad \quad \hspace{0.5em} & $\tau_d=m/\gamma_t$\\[8pt]
                        \hline
                        \hline
        \end{tabular}
        }
        \caption{Relevant time scales in active underdamped particles.}
        \label{table:timescales}
\end{table}

\paragraph*{Model.--}
To demonstrate our results in detail, 
let us now consider a generic model for active underdamped particles in 2D, each having an internal drive, represented by an effective 
self-propulsion force $\vec{F}_{\text{SP},i}=\gamma_t v_0 \vec{u}(\theta_i)$
where $\vec{u}(\theta_i)=(\cos \theta_i, \sin \theta_i)$ is the direction of self-propulsion. The particles have identical diameters $\sigma$, masses $m$ 
and moments of inertia $I$. They interact via an excluded-volume repulsive force $\vec{F}_i$ (see Supplemental Material). 
Their velocities $\vec{v}_i$ and orientations $\theta_i$ evolve as 
\begin{align}   
\begin{split} \label{EqLangevinEquations}
 m \frac{\diff \vec{v}_i}{\diff t} &= -\gamma_t \vec{v}_i + \vec{F}_i + \vec{F}_{\text{SP},i} + \sqrt{2 k_B T_b \gamma_t} \boldsymbol{\eta}_i (t), \\ 
 I \frac{\diff^2 \theta_i}{\diff t^2} &= -\gamma_r \frac{\diff \theta_i}{\diff t} + \sqrt{2k_B T_b \gamma_r} \xi_i(t),
\end{split}
\end{align}
where $\boldsymbol \eta_i$, $\xi_i$ represent Gaussian white noise of zero-mean unit variance, $T_b$ is the effective bath temperature 
and $\gamma_t,\gamma_r$ are translational and rotational drag coefficients, yielding diffusion coefficients
$D_{t,r}=k_BT_b/\gamma_{t,r}$. To understand the behavior of active underdamped particles, it is instructive to 
define three characteristic time scales (see table~\ref{table:timescales}): the persistence time $\tau_p=1/D_r$, 
after which the directed motion of active particles is randomized by rotational diffusion, 
the mean time between collisions for a given particle $\tau_c=\pi \sigma/(4v_0\varphi)$, where $\varphi=N\pi \sigma^2/(4L_x L_y)$ is the area fraction, and 
the inertial time scale $\tau_d=m/\gamma_t$, characterizing the time a particle at rest needs to reach its terminal speed.
(In principle, the moment of inertia $I$ leads to an additional timescale ($I/\gamma_r$), but it turns out to be 
largely irrelevant to our results and is thus kept constant to $I=0.33 \epsilon \tau_p^2$ (see Supplementary Material).

\begin{figure*}
\includegraphics[width=\textwidth]{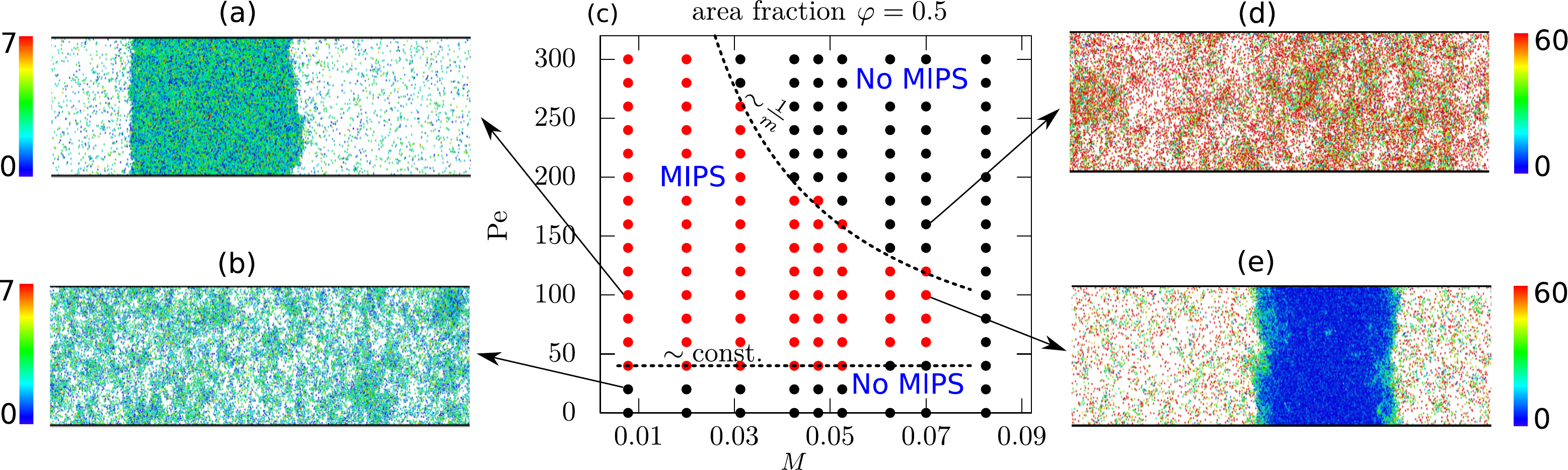}
\caption{\small Nonequilibrium phase diagram at $\varphi=0.5$ (c). Panels (a), (b), (d), and (e) 
represent snapshots from our simulations ($L_x \times L_y = 350\sigma \times 70\sigma$) at state points indicated in the phase diagram. Each simulation has been performed in a box of size $L_x \times L_y = 850\sigma \times 170\sigma$, comprising $N\approx 10^5$ particles. Colors represent kinetic energies of individual particles 
in units of $k_B T$. A hot-cold coexistence is visible in 
panel (e). Dashed lines in (c) show scaling predictions 
for the phase boundary between the homogeneous and phase-separated state.  
}
\label{fig:phasediagram}      
\end{figure*}
Fixing the area fraction to a regime where MIPS can occur ($\varphi=0.5$), 
the behavior of our system is mainly controlled by two parameters, which can be expressed as ratios of the 
relevant timescales: 
$M=\tau_d/\tau_p$, which is a reduced mass measuring 
the impact of inertia, and the P\'{e}clet number $\text{Pe}=v_0/(D_r \sigma) \propto \tau_p/(\tau_c \varphi)$ (see  Supplemental Material), 
measuring the strength of self-propulsion by
comparing ballistic to a diffusive motion.



\paragraph*{Nonequilibrium phase diagram.--}
To explore the impact of inertia on the collective behavior of active particles, we first explore the phase-diagram 
using large-scale simulations based on  
LAMMPS~\cite{Plimpton:1995}. 
If $M\rightarrow 0$, 
inertia plays no role and the particles are essentially overdamped. Accordingly, for 
$M \lesssim 10^{-4}$, we recover the usual behavior: at fixed area fraction $\varphi=0.5$, the particles undergo MIPS~\cite{Stenhammar2013,Redner2013} 
when the
P\'{e}clet number is large enough ($\text{Pe} \gtrsim 20$), leading to a dense liquid phase, coexisting with a gas phase (Fig.~\ref{fig:phasediagram}(a)), 
further characterized in Supplemental Material.  
For moderate inertia ($0.03 \le M \le 0.07$), we still require $\text{Pe}$ to exceed a certain threshold to allow the system to 
phase separate into a liquid and a coexisting gas (Fig.~\ref{fig:phasediagram}(e)). 
However, when further increasing Pe, 
strikingly, MIPS disappears and the system remains in the disordered phase (Fig.~\ref{fig:phasediagram}(d)). Thus, MIPS is reentrant for 
underdamped active particles. 
Finally, when inertia is even stronger $M \gtrsim 0.08$, MIPS does not occur at all. 
Overall, this leads to the phase diagram shown in 
Fig.~\ref{fig:phasediagram}(c).
The qualitative structure of this phase diagram can be understood based on simple scaling arguments. 
To see this, let us first remember how MIPS arises for overdamped particles:
consider a particle self-propelling towards a small dense cluster of particles; when colliding, the particle stops and is blocked by the cluster, until rotational 
diffusion turns its self-propulsion direction away from the cluster on a timescale $\tau_p=1/D_r$. 
When the time in between collisions $\tau_c$ is smaller than $\tau_p$, the rate of particles entering the cluster exceeds the leaving-rate and the 
cluster rapidly grows~\cite{Fily2012,Buttinoni2013}, later proceeding slowly towards phase separation. 
This criterion explains the existence of a (lower) critical P\'{e}clet number. 
Since both $\tau_c, \tau_p$ are mass-independent, we expect the lower critical \text{Pe} number also to be mass-independent: 
\begin{equation}
\tau_p \gtrsim \tau_c \quad \Rightarrow \text{Pe}_{1} = {\rm const.}
\end{equation}
as approximately observed in Fig.~\ref{fig:phasediagram}(c).
To understand the upper critical \text{Pe} number, note that MIPS requires a localized slow-down of particles to occur. 
Thus, at very high collision rates (due to high \text{Pe}), underdamped particles bounce back multiple times on the inertial time scale $\tau_d$, and can therefore not slow down locally.
We, therefore, expect that MIPS occurs only if
\begin{equation}
\tau_c \gtrsim \tau_d \quad \Rightarrow \text{Pe}_{2} \propto 1/m,
\end{equation} 
which yields the scaling law $\text{Pe}\sim 1/m$ shown as the upper dashed line in Fig.~\ref{fig:phasediagram}(c) and corresponds to our 
simulation results.

\begin{figure*}
\includegraphics[width=\linewidth]{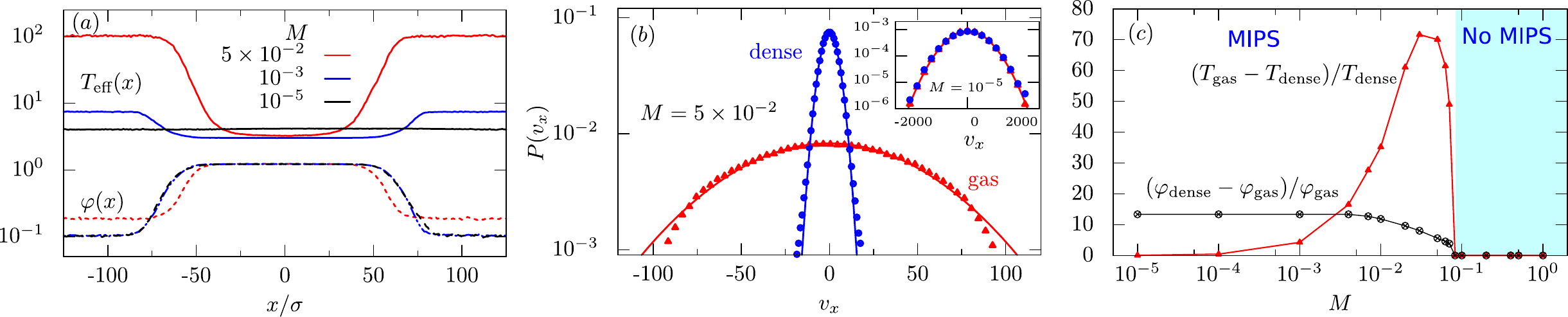}
\caption{\small
(a) Spatial profiles of the effective temperature $T_\text{eff}(x)+2.0$ (solid lines) and local area fraction $\varphi(x)$ (dashed lines) for 
different reduced masses $M$. (b) Steady-state distributions of particle velocities $v_x$ for 
moderate inertia $M=5\times 10^{-2}$. 
Solid lines are fits to the Maxwell-Boltzmann distribution $P(v_x)=\sqrt{m/(2 \pi T_\text{eff} )} \exp[-mv_x^2/(2T_\text{eff})]$, where $T_\text{eff}$ 
is the kinetic temperature. Inset: $P(v_x)$ for vanishing inertia $M=10^{-5}$. 
(c) The relative temperature and area fraction difference between the two phases as a function of inertia. Other parameters: $\text{Pe}=100$, $\varphi=0.5$.}
\label{fig:effectiveT}
\end{figure*}

\paragraph*{Temperature difference in coexisting phases.--}
Let us now explore the properties of the resulting liquid and the coexisting gas, in parameter regimes where MIPS takes 
place. 
While in the overdamped case ($M\rightarrow 0$), particles in the liquid and in the coexisting gas are equally fast on average 
as shown by the colors in Fig.~\ref{fig:phasediagram}(a), this changes dramatically when inertia becomes significant. 
Following the colors in Fig.~\ref{fig:phasediagram}(e) we see, strikingly, that particles in the liquid (blue dots) are much slower than in the gas 
(green, yellow and red dots).
Before discussing the origin of this remarkable temperature difference, let us quantify it more detail. To this end, we 
define the kinetic temperature as $T_\text{eff}(x)=\frac{1}{2}m\langle v^2(x) \rangle$, which is the 
kinetic energy per particle, averaged 
along the lateral coordinate. As shown in Fig.~\ref{fig:effectiveT}(a), $T_\text{eff}$ is uniform for $M=10^{-5}$, but develops 
a massively nonuniform shape when increasing $M$ to $0.05$ (see Supplementary Movies S1 and S2, respectively). 
Fig.~\ref{fig:effectiveT}(c) quantifies the resulting 
temperature difference, showing $(T_\text{gas} -T_\text{dense})/T_\text{dense}$ as a function of $M$. 
Here, we see that the temperature in the dilute phase can be almost two orders of magnitude larger than in the dense phase. 
(Note that Fig.~\ref{fig:effectiveT}(c) shows that the temperature difference has a maximum at some $M$ value before MIPS disappears, 
and then decreases again; 
this is probably a consequence of the fact, that the collision rate in the gas phase increases in the corresponding parameter domain, 
which cools the gas, as we will see below.)
This is further reflected by the velocity distribution $P(v_x)$ in Fig.~\ref{fig:effectiveT}(b), showing a far-broader 
distribution for the gas phase than for the dense one, but only if inertia is significant (see inset). 

\begin{figure}
	\includegraphics[width=\linewidth]{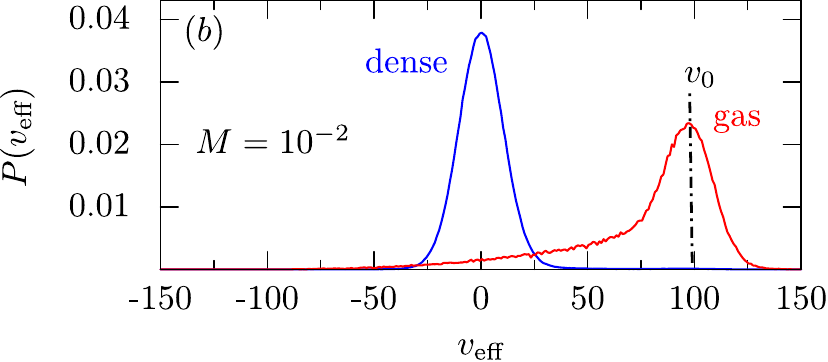}
	\caption{\label{fig:EnergyDissip}
Probability distributions of effective speeds in the gas phase as well as in the dense phase.
Other parameters: $\text{Pe}=100$, $\varphi=0.5$.
	}
\end{figure}

\paragraph*{Power-balance.--}
To understand the temperature difference quantitatively, we now derive a power-balance equation. 
Multiplying the translational part of Eq.~\eqref{EqLangevinEquations} by $\vec{v}$, 
and averaging over all particles in a given phase, we obtain

\begin{equation}
\label{eq:kenerg1}
\begin{split}
\frac{1}{2} m\frac{\diff \langle v^2(t) \rangle}{\diff t} &= -\gamma_t \langle v^2(t) \rangle + \langle \vec{v}(t) \cdot \vec{F}(t) \rangle +  
\langle \vec{v}(t) \cdot \vec{F}_\text{SP}(t) \rangle \\
 &+ \sqrt{2 k_B T_b \gamma_t}\langle \vec{v}(t) \cdot \boldsymbol{\eta}(t) \rangle.
\end{split}
\end{equation}
Here, the left hand side equals the time 
derivative of the effective temperature $\partial T_\text{eff}/\partial t$; $\gamma_t \langle v^2(t) \rangle=2T_\text{eff}/\tau_d$ 
describes the energy dissipation rate due to Stokes drag 
and 
$\langle \vec{v}(t) \cdot \vec{F}(t) \rangle$ represents the dissipated power due to interactions among the particles, which is negligible here since 
particle collisions are elastic, see Supplementary Fig. S4.
The third-term $\langle \vec{v}(t) \cdot \vec{F}_\text{SP}(t) \rangle$ represents the self-propulsion power.
The last-term is related to the bath temperature by the following relation $\sqrt{2 k_B T_b \gamma_t}\langle \vec{v}(t) \cdot \boldsymbol{\eta}(t) \rangle=2k_B T_b \gamma_t/m=2k_BT_b/\tau_d$, 
which is identical in the gas and in the dense phase. 
Plugging these expressions into Eq.~\eqref{eq:kenerg1}, and 
and using that $\partial T_\text{eff}/\partial t=0$ in each phase individually in steady state, we obtain
%
\begin{equation}
\label{eq:coexisting}
T_{\text{gas}} - T_{\text{dense}} = \frac{\tau_d}{2}\big[ \langle \vec{v} \cdot \vec{F}_\text{SP} \rangle_{\text{gas}} - 
\langle \vec{v} \cdot \vec{F}_\text{SP} \rangle_\text{dense} \big].
\end{equation}
Therefore, if and only if $\tau_d \neq 0$, self-propulsion can create a temperature difference in coexisting phases.
Since $\tau_d = 0$, in overdamped particles, both phases have the same kinetic temperature. 
In contrast, for underdamped particles we have $\tau_d \neq 0$.
The contributions of the individual terms to the power balance is visualized in Supplementary Fig. S4, revealing that the 
self-propulsion power is much higher in the gas phase than in the dense phase and dominates the kinetic temperature (rather than diffusion as for overdamped particles).
To see, why the self-propulsion power is different in the gas phase compared to the dense phase, we explore the distribution of 
the particle effective speeds $v_\text{eff} =\vec{v} \cdot \vec{u}$ 
in both phases; here
$\langle \vec{v} \cdot \vec{F}_\text{SP} \rangle=\gamma_t v_0 \langle v_\text{eff} \rangle $. 
Thus, Figure~\ref{fig:EnergyDissip} shows that the average effective speed in the gas phase is $v_0$, whereas negative speed values are rare, showing that particles in the gas phase rarely move against their 
self-propulsion direction (Fig.~\ref{fig:hot_and_cold_cartoon}, left panel). 
This suggests that $\langle \vec{v} \cdot \vec{F}_\text{SP} \rangle_{\text{gas}} \sim \gamma_t v_0^2$. In contrast, in the dense phase, 
the effective particle speed is almost symmetrically distributed around 0, which results 
from the fact that particles have no space to move and bounce back after each collision; thus, they move against their self-propulsion direction 
about half of the time (Fig.~\ref{fig:hot_and_cold_cartoon}, right panel), which implies $\langle \vec{v} \cdot \vec{F}_\text{SP} \rangle_{\text{dense}} \sim 0$.

\paragraph*{Conclusion.--}
Unlike equilibrium systems, self-driven active particles can self-organize into a 
liquid and a coexisting gas phase at different temperatures. 
This result exemplifies a route to use self-driven particles to create a self-sustained temperature gradient, which might serve, in principle, 
as a novel paradigm to create isolating layers at the microscale, e.g. to keep bodies at different temperatures. 

On a more fundamental level, our results show that motility-induced phase separation, one of the best explored 
phenomenon in active matter research, is fundamentally different from a liquid-gas phase separation -- an insight which has been 
curtained by the focus on overdamped particles so far. 
As a consequence, the phenomenology of 
motility-induced phase separation is even richer than anticipated previously - it can, in particular, lead to 
phenomena at the macroscale which are fundamentally beyond equilibrium physics.

For future studies, it would also be interesting to study the effect of inertia on anisotropic active particles~\cite{Nguyen:2014,Farhadi:2018,kokot:2017,Aubret:2018} where translational and rotational motions are coupled. Specifically for such particles, ref.~\cite{Petrelli:2018}
has recently observed (but hardly analyzed) the occurrence of different kinetic energies in coexisting phases, suggesting that the present findings survive for particles of nonspherical shape. 

An interesting challenge would also be to derive a 
microscopic
theory for motility-induced phase separation in underdamped particles to predict the joint temperature and density
profiles across the interface between the two coexisting states~\cite{Miyazaki:2018}. Such an approach needs to be designed for non-isothermal situations as considered recently in Enskog kinetic
theories \cite{Brey:2015,Grazo:2018}
or in dynamical density functional theory
\cite{Wittkowski:2012,Anero:2013}.

\begin{acknowledgments}
We thank Christian Scholz and Alexei Ivlev for fruitful discussions. This work is supported by the German Research Foundation (Grant No. LO 418/23-1)
\end{acknowledgments}

%

\onecolumngrid
\clearpage
\section{Supplementary Material}

\section*{Simulations}
Simulations are performed with a slightly modified version of LAMMPS~\cite{Plimpton:1995}, which integrates the equations of motion given in Eq.~(1) using the Euler method. The conservative force on particle $i$ from particle $j$ is $\vec{F}_i=-\vec{\nabla}_i u(r_{ij})$, which results from a purely repulsive WCA potential~\cite{Weeks:1971}:  
\begin{align*}
u(r_{ij})=
\begin{cases}
4\epsilon \big [ (\frac{\sigma}{r_{ij}})^{12} - (\frac{\sigma}{r_{ij}})^{6} \big ] + \epsilon ,& r_{ij}/\sigma \le 2^{1/6}\\
\quad  0, & r_{ij}/\sigma \textgreater 2^{1/6}
\end{cases}
\end{align*}
where $\epsilon=k_BT$ is the interaction strength, and $r_{ij}$ is the distance between particles $i$ and $j$. The equations of motion are integrated with a time step $\delta t=10^{-5}\tau_p$. Recent experiments~\cite{Scholz2018b} on microflyers reveal that diffusion coefficients ($D_r$ and $D_t$) and friction coefficients ($\gamma_r$ and $\gamma_t$) are not related by the Stokes-Einstein relation. Thus, for simplicity, we choose $\gamma_t=\gamma_r/\sigma^2$ as shown, e.g., in Ref.~\cite{Lei:2019}. 

\begin{figure}[h]
	\includegraphics[width=0.5\linewidth]{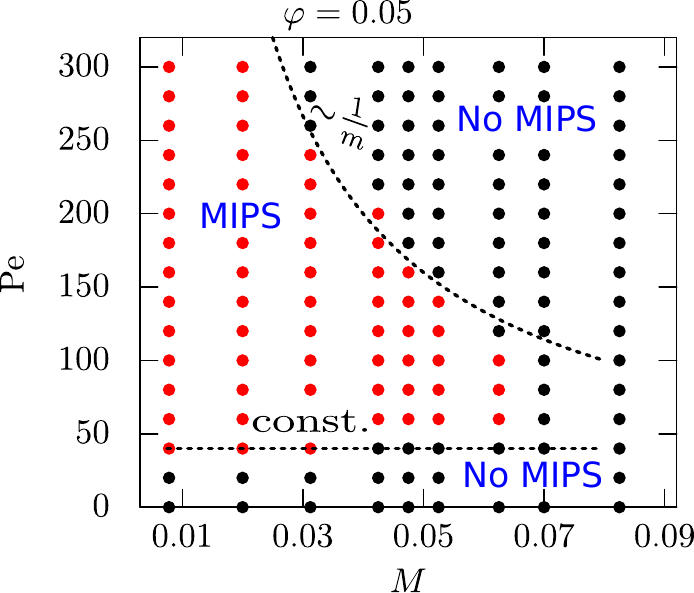}
	\caption{\label{fig:additionalphasediagram}
		Nonequilibrium phase diagram same as Fig. 2(c), but now for $I=0.066\epsilon \tau_p^2$. 		
	}
\end{figure}

In order to clarify the importance of the moment of inertia $I$, we have performed simulations with two different moments of inertia $I=0.33\epsilon \tau_p^2$ (Fig.~2(c) in the main text) and $I=0.066\epsilon \tau_p^2$ (Supplementary Fig.~\ref{fig:additionalphasediagram}). These two figures display qualitatively similar results, which implies that we are close to overdamped rotational dynamics, where $I=0$.


 
\section*{Bouncing ball picture}
\begin{figure}
\includegraphics[width=0.5\linewidth]{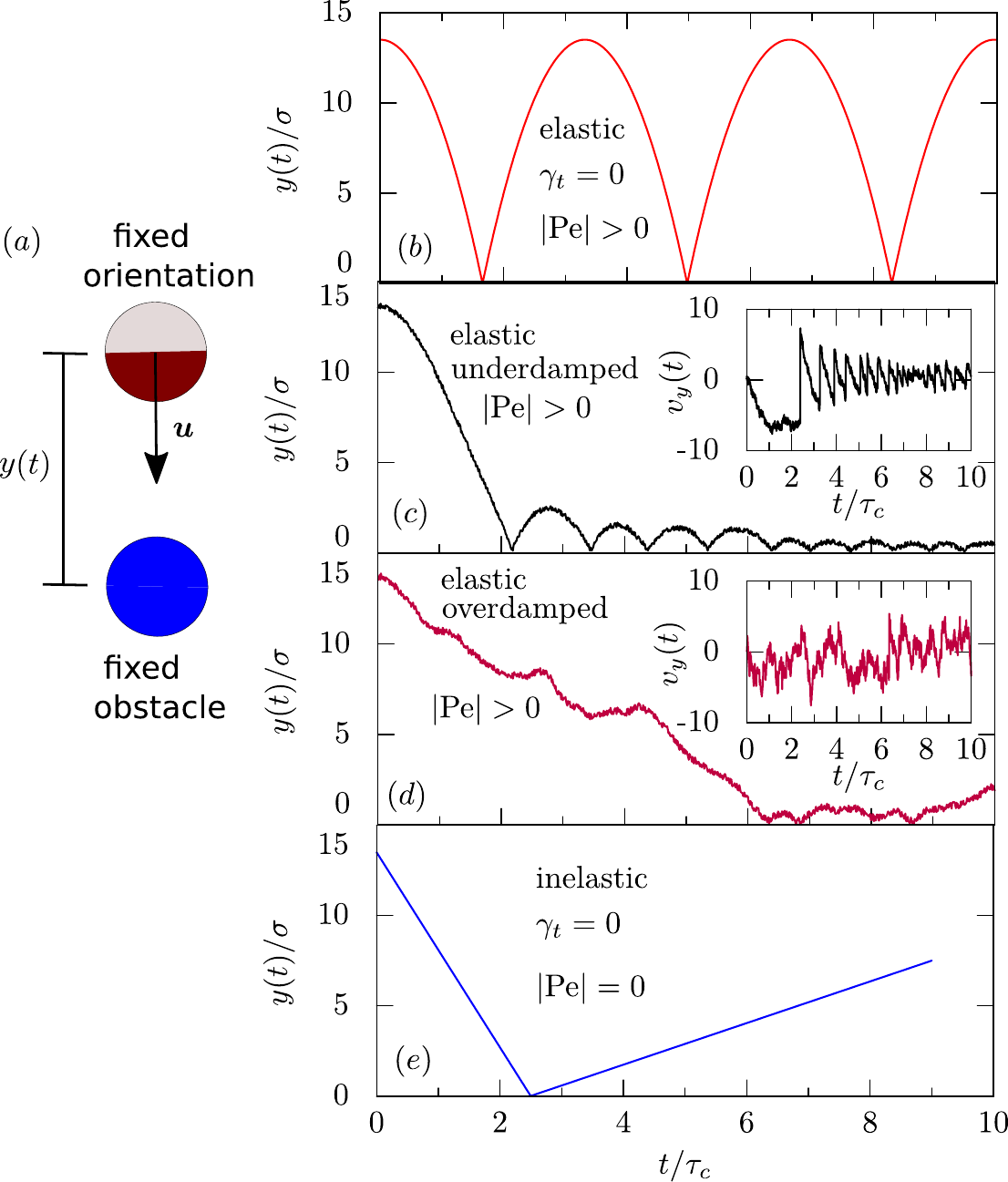}
\caption{\label{fig:bouncing ball}
(a) A ball with fixed orientation bounces back elastically from a fixed obstacle (blue particle). Typical trajectories (and velocities in insets) of a bouncing ball for (b) 
vanishing drag $\gamma_t=0$ , (c) finite drag $\gamma_t \neq 0$ (underdamped), and (d) infinite drag $\gamma_t=\infty$ (overdamped). 
(e) A typical trajectory of a ball when it encounters an inelastic collision rather than drag. 
}
\end{figure}

To develop an intuition for the emergence of temperature differences let us exploit a simple formal analogy: 
the dynamics of an active particle with 
fixed orientation, which is elastically reflected by a fixed obstacle, is identical to the dynamics 
of an elastically bouncing ball under the influence of gravity (representing 
self-propulsion) and Stokes drag 
(Fig.~\ref{fig:bouncing ball}(a)). 
To characterize the bouncing dynamics, we show the vertical position $y(t)$ 
as a function of time $t$ in Fig.~\ref{fig:bouncing ball}. For vanishing drag, 
$\gamma_t=0$, energy is conserved and the ball bounces periodically without slowing down (Fig.~\ref{fig:bouncing ball}(b)). 
However, when experiencing drag, 
the ball, initially at rest, accelerates due to gravity to a velocity which cannot exceed $v_0$
before hitting the 
fixed obstacle (Fig.~\ref{fig:bouncing ball}(c) 
and inset). 
The ball bounces back elastically, preserving its speed upon the collision, 
but now ascends against the gravitational force to a turning point below the starting position.
From here, the ball accelerates towards the obstacle again, but has less space to accelerate this time. Thus, each time the ball hits the obstacle, 
it is slower.  
The same slow-down mechanism applies to a particle entering the dense phase and encountering a series of collisions, each time 
bouncing back, against its self-propulsion direction, and having less space to accelerate.
This is in stark contrast to the behavior in the gas phase, where collisions are rare and particles have enough time to reach their 
terminal speed $v_0$ in between collisions. Thus, the active gas is much 'hotter' than the active liquid. 
The behavior of an overdamped bouncing ball is yet different (Fig.~\ref{fig:bouncing ball}(d) and inset): this ball reaches its terminal speed instantaneously; 
when hitting the obstacle, it does not 
bounce back, and does not move any further, apart from translational diffusion. 
Here, while directed motion immediately stops when hitting the obstacle, 
the actual velocity of the particle hardly changes: This is because the instantaneous speed of overdamped particles is dominated by the diffusive micromotion, not 
by self-propulsion. Consequently, overdamped particles are equally fast in the gas and in the liquid, yielding identical temperatures in both phases -- as in 
equilibrium. 
Finally, to contrast the present slow-down mechanism, crucially based on self-propulsion, from the scenario in vibrated granular particles, 
let us emphasize that 
the latter corresponds to a ball experiencing inelastic collisions, i.e. to a case where kinetic energy is drained from the 
system upon a collision. 

\section*{Nonequilibrium phase diagram}
To construct the phase diagram, an elongated box with periodic boundary conditions is used. Simulations were run up to $10^5 \tau_p$ in order to reach the steady state. To characterize the phase-separation, we measure the distribution $P(\varphi_\text{loc})$ of the local free-area $\varphi_\text{loc}$ of active underdamped particles using the Voronoi tessellation method~\cite{Rycroft:2009} (see Supplementary Fig.~\ref{fig:freevolume}). Once the free-area distribution is bimodal, we identify it as a phase-separated state. 

\section*{Area fraction difference in coexisting phases}
In the phase-separated state, we can measure the local area fractions in the two different phases by dividing the simulation 
box into slabs of width $\simeq 0.5\sigma$. 
We find that the area fraction profiles (dashed lines) in Fig.~3(a) in the main text
are similar to ABPs and can be fitted to a hyperbolic tangent function
\begin{equation}
\label{eq:densityprofile}
\varphi(x)=\frac{1}{2}(\varphi_\text{dense} + \varphi_\text{gas}) - \frac{1}{2}(\varphi_\text{dense} - \varphi_\text{gas}) \tanh \Big [ \frac{2(x-x_0)}{w} \Big],
\end{equation}
where $x_0$ and $w$ are the location and width of the gas-liquid interface. 
We extract the corresponding area fractions of the gas phase $\varphi_\text{gas}$ and 
the dense phase $\varphi_\text{dense}$ by fitting each side of the interface using Eq.~\eqref{eq:densityprofile}.
In Fig.~3(c) (in the main text), we plot the relative area fraction difference 
$(\varphi_{\text{dense}} -\varphi_{\text{gas}})/\varphi_{\text{}gas}$ in coexisting phases 
by varying inertia while keeping the P\'{e}clet number fixed  at $\text{Pe}=100$. 
Notably we find that the area fraction difference between the two phases is 10 times higher than 
the gas phase and interfacial width $w \simeq 20\sigma$. As we move from phase-separated to a homogeneous state with increasing inertia $M$ at fixed $\text{Pe}$, the relative area fraction decreases monotonically towards a critical inertia $M\approx 0.08$. This behavior is similar to the first-order-phase transition, but occurs in a non-equilibrium setup. Most importantly, the control parameter is inertia $M$ instead of the thermodynamic temperature.

\begin{figure}[h]
	\includegraphics[width=0.5\linewidth]{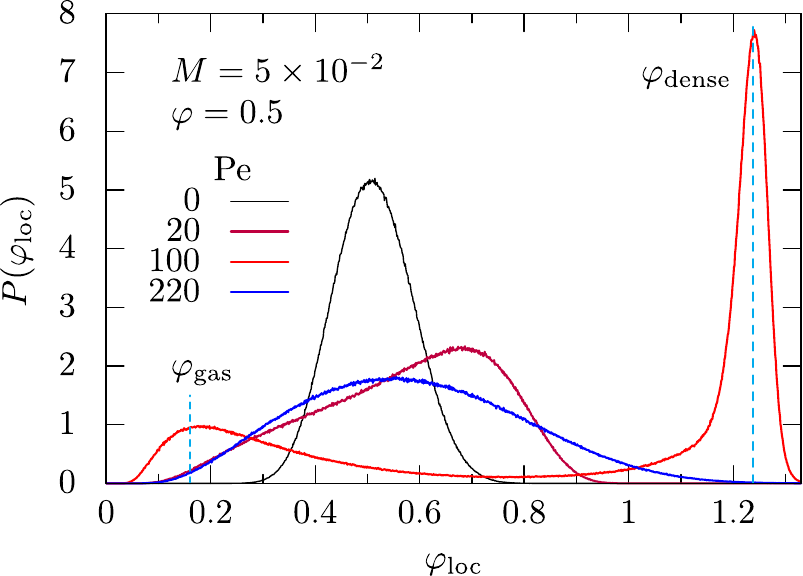}
	\caption{\label{fig:freevolume}
		Local free-area distributions for various P\'{e}clet numbers at fixed inertia $M=5\times 10^{-2}$ and global area fraction $\varphi=0.5$. The distribution is peaked around the overall area fraction in equilibrium ($\text{Pe}=0$). It broadens near the critical P\'{e}clet number $\text{Pe}=20$. For $\text{Pe}=100$, the distribution becomes bimodal as the system phase separates into a gas and a dense phase. As \text{Pe} is increased further, it becomes unimodal (homogeneous) again.
	}
\end{figure}

\begin{figure}[h]
	\includegraphics[width=0.5\linewidth]{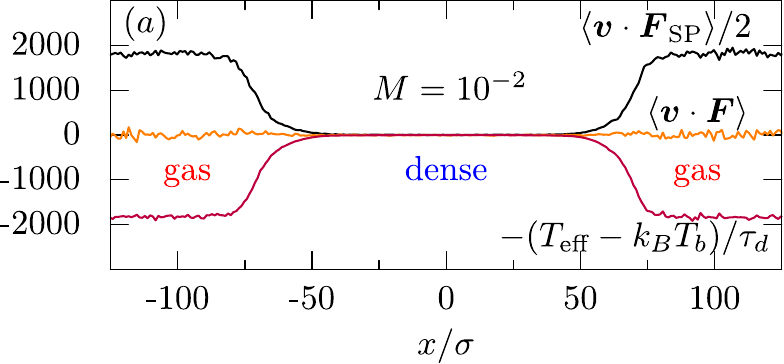}
	\caption{\label{fig:powerbalance}
		 Power balance in the phase-separated state, 
where the injected power by active forces is balanced by energy dissipation rate due to Stokes drag. 
	}
\end{figure}

\section*{Legends to movies}
In all movies $\varphi=0.5$, $L_x \times L_y=850\sigma \times 170\sigma$, $N=10^{5}$, $\text{Pe}=100$, while reduced mass $M$ and reduced temperature $T^{*}=T_\text{eff}/k_BT$ are provided for each movie. The simulation time $t$ is measured in units of $\tau_p$.

\begin{enumerate}
	\item \textbf{Movie S1:} Underdamped active particles with $M=10^{-5}$.
	    Here, coexisting phases have the same temperature as shown by the colors, just like in equilibrium physics.
	\item \textbf{Movie S2:} Underdamped active particles with $M=0.05$. A massive temperature difference emerges between the two phases. In particular, particles in the dense phase (blue dots) are 'colder' than in the gas phase (green, yellow and red dots).
\end{enumerate}

\end{document}